\documentclass[letterpaper, 10pt, conference]{ieeeconf}      

\IEEEoverridecommandlockouts                              
\usepackage{epsfig}
\usepackage{amssymb}
\usepackage{graphicx}
\usepackage{graphics}
\usepackage{amsmath}

\title{\LARGE \bf Efficient Energy Management Policies for Networks with Energy Harvesting Sensor Nodes \thanks{This work was partially supported by a research grant from Boeing Corporation.}}

\author{Vinod Sharma, Utpal Mukherji and Vinay Joseph  
\thanks{Vinod Sharma, Utpal Mukherji, Vinay Joseph are with the Dept of Electrical Communication Engineering, Indian Institute of Science, Bangalore, India. Email: {\tt  \{ vinod,utpal,vinay \}@ece.iisc.ernet.in}}
\thanks{This work has been submitted
to the IEEE for publication. Copyright may be transferred
without notice, after which this version may no longer be accessible.}}

\begin{document}

\maketitle
\thispagestyle{empty}
\pagestyle{empty}

\begin{abstract}
We study sensor networks with energy harvesting nodes. The generated energy at a node can be stored in a buffer. A sensor node periodically senses a random field and generates a packet. These packets are stored in a queue and transmitted using the energy available at that time at the node. For such networks we develop efficient energy management policies. First, for a single node, we obtain policies that are throughput optimal, i.e., the data queue stays stable for the largest possible data rate. Next we obtain energy management policies which minimize the mean delay in the queue. We also compare performance of several easily implementable sub-optimal policies. A greedy policy is identified which, in low SNR regime, is throughput optimal and also minimizes mean delay. Next using the results for a single node, we develop efficient MAC policies. 

\end{abstract}
\noindent
\textbf{Keywords:} Optimal energy management policies, energy harvesting, sensor networks, MAC protocols.

\section{Introduction}
\label{intro} 
Sensor networks consist of a large number of small, inexpensive sensor nodes. These nodes have small batteries with limited power and also have limited computational power and storage space. When the battery of a node is exhausted, it is not replaced and the node dies. When sufficient number of nodes die, the network may not be able to perform its designated task. Thus the life time of a network is an important characteristic of a sensor network and it is tied up with the life time of a node. 

Various studies have been conducted to increase the life time of the battery of a node by reducing the energy intensive tasks, e.g., reducing the number of bits to transmit (\cite{baek}, \cite{pradhan}), making a node to go into power saving modes: (sleep/listen) periodically (\cite{sinha}), using energy efficient routing (\cite{ratnaraj}, \cite{woo}) and MAC (\cite{ye}). Studies that estimate the life time of a sensor network include \cite{ratnaraj}. A general survey on sensor networks is \cite{akyildiz} which provides many more references on these issues.

In this paper we focus on increasing the life time of the battery itself by energy harvesting techniques (\cite{kansal}, \cite{niyato}). Common energy harvesting devices are solar cells, wind turbines and  piezo-electric cells, which extract energy from the environment. Among these, solar harvesting energy through photo-voltaic effect seems to have emerged as a technology of choice for many sensor nodes (\cite{niyato}, \cite{raghunathan}). Unlike for a battery operated sensor node, now there is potentially an \textit{infinite} amount of energy available to the node. Hence energy  conservation need not be the dominant theme. Rather, the issues involved in a node with an energy harvesting source can be quite different. The source of energy and the energy harvesting device may be such that the energy cannot be generated at all times (e.g., a solar cell). However one may want to use the sensor nodes at such times also. Furthermore the rate of generation of energy can be limited. Thus one may want to match the energy generation profile of the harvesting source with the energy consumption profile of the sensor node. It should be done in such a way that the node can perform satisfactorily for a long time, i.e.,  at least energy starvation should  not be the reason for the node to die. Furthermore, in a sensor network, the MAC protocol, routing and relaying of data through the network may need to be suitably modified to match the energy generation profiles of different nodes, which may vary with the nodes.

In the following we survey the literature on sensor networks with energy harvesting nodes. Early papers on energy harvesting in sensor networks are \cite{kansal1} and \cite{rahimi}. A practical solar energy harvesting sensor node prototype is described in \cite{jiang}. A good recent contribution is \cite{kansal}. It provides various deterministic theoretical models for energy generation and energy consumption profiles (based on $(\sigma, \rho)$ traffic models and provides conditions for \emph{energy neutral operation}, ie., when the node can operate indefinitely.  In \cite{jaggi} a sensor node is considered which is sensing certain interesting events. The authors study optimal sleep-wake cycles such that the event detection probability is maximized. A recent survey is \cite{niyato} which also provides an optimal sleep-wake cycle for solar cells so as to obtain QoS for a sensor node. 

MAC protocols for sensor networks are studied in \cite{awoo}, \cite{ye} and \cite{zhao}. A general survey is available in \cite{akyildiz} and \cite{kredo}. Throughput optimal opportunistic MAC protocols are discussed in \cite{lin}.

In this paper we summarize our recent results (\cite{joseph}), on sensor networks with energy harvesting nodes and based on them propose new schemes for scheduling a MAC for such networks. The motivating application is estimation of a random field which is one of the canonical applications of sensor networks. The above mentioned theoretical studies are motivated by other applications of sensor networks. In our application, the sensor nodes sense the random field periodically. After sensing, a node generates a packet (possibly after efficient compression). This packet needs to be transmitted to  a central node, possibly via other sensor nodes. In an energy harvesting node, sometimes there may not be sufficient energy to transmit the generated packets (or even sense) at regular intervals and then the node may need to store the packets till they are transmitted. The energy generated can be stored (possibly in a finite storage) for later use.  

Initially we will assume that most of the energy is consumed in transmission only. We will relax this assumption later on. We find conditions for energy neutral operation of a node, i.e., when the node can work forever and its data queue is stable. We will obtain policies which can support maximum possible data rate. 

We also obtain energy management (power control) policies for transmission which minimize the mean delay of the packets in the queue.

We use the above energy mangement policies to develop channel sharing policies at a MAC (Multiple Access Channel) used by energy harvesting sensor nodes.   

We are currently investigating appropriate routing algorithms for a network of energy harvesting sensor nodes.

The paper is organized as follows. Section \ref{model} describes the model for a single node and provides the assumptions made for data and energy generation. Section \ref{stability} provides conditions for energy neutral operation of a node. We obtain stable, power control policies which are throughput optimal. Section \ref{opt} obtains the power control policies which minimize the mean delay via Markov decision theory. A greedy policy is shown to be throughput optimal and provides minimum mean delays for linear transmission. Section \ref{general} provides a throughput optimal policy when the energy consumed in sensing and processing is non-negligible. A sensor node with a fading channel is also considered.  Section \ref{simulation} provides simulation results to confirm our theoretical findings and compares various energy management policies.  Section \ref{tree} introduces a multiple access channel (MAC) for energy harvesting nodes. Section \ref{orthogonal} provides efficient energy management schemes for orthogonal MAC protocols. Sections \ref{opportunistic} and \ref{csma} consider MACs with fading channels (orthogonal and CSMA respectively). Section \ref{msimulation} compares the different MAC policies via simulations. Section \ref{conclude} concludes the paper. 

\section{Model for a single node}
\label{model} In this section we present our model for a single energy harvesting sensor node.

\begin{figure}[!ht]
\begin{center}
\includegraphics[scale=0.43]{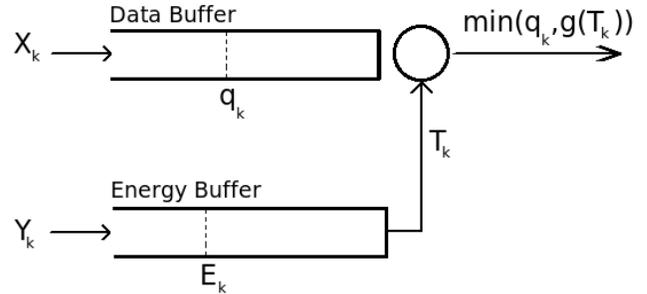}
\caption{The model} \label{fig1}
\end{center}
\end{figure}

We consider a sensor node (Fig. \ref{fig1}) which is sensing a random field and generating packets to be transmitted to a central node via a network of sensor nodes. The system is slotted. During slot $k$ (defined as time interval $[k, k+1]$, i.e., a slot is a unit of time) $X_k$ bits are generated by the sensor node. Although the sensor node may generate data as packets, we will allow arbitrary fragmentation of packets during transmission. Thus, packet boundaries are not important and we consider bit strings (or just fluid). The bits $X_k$ are eligible for transmission in $(k+1)$st slot. The queue length (in bits) at time $k$ is $q_k$. The sensor node is able to transmit $g(T_k)$ bits in slot $k$ if it uses energy $T_k$. We assume that transmission consumes most of the energy in a sensor node and ignore other causes of energy consumption (this is true for many low quality, low rate sensor nodes (\cite{raghunathan})). This assumption will be removed in Section \ref{general}. We denote by $E_k$ the energy available in the node at time $k$. The sensor node is able to replenish energy by $Y_k$ in slot $k$. 

We will initially assume that $ \{ X_k \}$ and $ \{ Y_k \}$ are \emph{iid} (independent, identically distributed) but will generalize this assumption later. It is important to generalize this assumption to capture realistic traffic streams and energy generation profiles. 

The processes $ \{ q_k \}$ and $ \{ E_k \}$ satisfy 
\begin{eqnarray}
\label{eqn1}
q_{k+1} & = & (q_k - g(T_k))^{+} + X_k , \\
E_{k+1} & = & (E_k - T_k) + Y_k. \label{eqn2}
\end{eqnarray} where $T_k \leq E_k$. This assumes that the data buffer and the energy storage buffer are infinite. If in practice these buffers are large enough, this is a good approximation. If not, even then the results obtained provide important insights and the policies obtained often perform well for the finite buffer case. 

The function $g$ will be assumed to be monotonically non-decreasing. An important such function is given by Shannon's capacity formula \begin{equation} g(T_k) = \frac{1}{2} log (1 + \beta T_k) \label{allert3} \end{equation} for Gaussian channels where $\beta$ is a constant such that $ \beta \ T_k$ is the SNR. This is a non-decreasing concave function. At low values of $T_k$, $ g(T_k) \sim \beta_1 \ T_k$, i.e., $g$ becomes a linear function. Since sensor nodes are energy constrained, this is a practically important case. Thus in the following we limit our attention to linear and concave nondecreasing functions $g$. We will also assume that $g(0) =0$ which always holds in practice.

Many of our results (especially the stability results) will be valid when $\{ X_k \}$ and $\{ Y_k \}$ are stationary, ergodic. These assumptions are general enough to cover most of the stochastic models developed for traffic (e.g., Markov modulated) and energy harvesting. 

Of course, in practice, statistics of the traffic and energy harvesting models will be time varying (e.g., solar cell energy harvesting will depend on the time of day). But often they can be approximated by piecewise stationary processes. For example, energy harvesting by solar cells could be taken as being stationary over one  hour periods. Then our results could be used over these time periods. Often these periods are long enough for the system to attain (approximate) stationarity and for our results to remain meaningful.

In Section \ref{stability} we study the stability of this queue and identify easily implementable energy management policies which provide good performance.

\section{Stability}
 \label{stability} 
We will obtain a necessary condition for stability. Then we present a transmission policy which achieves the necessary condition, i.e., the policy is throughput optimal. The mean delay for this policy  is not minimal. Thus, we obtain other policies which provide lower mean delay. In the next section we will consider policies which minimize mean delay.

The proofs of Lemma 1 and Theorems 1-4 are provided in \cite{joseph}. 

Let us assume that we have obtained an (asymptotically) stationary and ergodic transmission policy $\{ T_k \}$ which makes $\{ q_k \}$ (asymptotically) stationary with the limiting distribution independent of $q_0$. Taking $\{T_k \}$ asymptotically stationary seems to be a natural requirement to obtain (asymptotic) stationarity of $\{ q_k \}$.

In the following $X, T, Y$ will denote generic r.v.s with the distributions of $X_1, T_1, Y_1$ respectively. 

\vspace{0.1cm}
\noindent
{\bf Lemma 1} Let $g$ be concave nondeceasing and $\{X_k \}, \{Y_k \}$ be stationary, ergodic sequences. For $\{ T_k \}$ to be an asymptotically stationary, ergodic energy management policy that makes $\{q_k \}$ asymptotically stationary with a proper stationary distribution $\pi$, it is necessary that $E [X] < E_{\pi} [g(T)] \leq g(E[Y]).$ \hspace*{6.7cm} $ \blacksquare$

Let \begin{equation} T_k = min(E_k, E[Y]-\epsilon) \label{disp} \end {equation} where $\epsilon$ is an appropriately chosen small constant (see statement of Theorem 1). We show that it is a throughput optimal policy, i.e., using this $T_k$ with $g$ satisfying the assumptions in Lemma 1, $\{ q_k \}$ is asymptotically stationary and ergodic. 
\vspace{0.12cm}

\noindent
{\bf Theorem 1} If $\{ X_k \}, \{ Y_k \}$ are stationary, ergodic and $g$ is continuous, nondecreasing, concave then if $E[X] < g(E[Y])$, (\ref{disp}) makes the queue stable (with $\epsilon > 0$ such that $ E[X] < g(E[Y] - \epsilon)$), i.e., it has a unique, stationary, ergodic distribution and starting from any initial distribution, $q_k$ converges in total variation to the stationary distribution. 
\hspace*{8.5cm} $ \blacksquare$

Henceforth we denote the policy (\ref{disp}) by TO.

From results on GI/GI/1 queues (\cite{asmussen}), if $\{ X_k \}$ are \emph{iid}, $E[X] < g(E[Y]), T_k =min(E_k, E[Y] - \epsilon)$ and $E[X^{\alpha}] < \infty $ for some $ \alpha > 1$ then the stationary solution $ \{ q_k \}$ of (\ref{eqn1}) satisfies $E[q^{\alpha-1}] < \infty$.

Taking $T_k = Y_k$ for all $k$ will provide stability of the queue if $E[X] < E[g(Y)]$. If $g$ is linear then this coincides with the necessary condition. If $g$ is strictly concave then $E[g(Y)] < g(E[Y]) $ unless $Y \equiv E[Y]$. Thus $T_k = Y_k$ provides a strictly smaller stability region. We will be forced to use this policy if there is no buffer to store the energy harvested. This shows that storing energy allows us to have a larger stability region. We will see in Section \ref{simulation} that storing energy can also provide lower mean delays.

Although TO is a throughput optimal policy, if $q_k$ is small, we may be wasting some energy. Thus, it appears that this policy does not minimize mean delay. It is useful to look for policies which minimize mean delay. Based on our experience in \cite{sharma}, the Greedy policy

\begin{equation}
 T_k = min(E_k, f(q_k)) \label{eqn5}
\end{equation} where $f = g^{-1}$, looks promising. In Theorem 2, we will show that the stability condition for this policy is $E[X] < E[g(Y)] $ which is optimal for linear $g$ but strictly suboptimal for a  strictly concave $g$ (just as the policy $T_k = Y_k$ discussed above). We will show in Section \ref{opt} that when $g$ is linear, (\ref{eqn5}) is not only throughput optimal, it also minimizes long term mean delay.

For concave $g$, we will show via simulations that (\ref{eqn5}) provides less mean delay than TO at low load. However since its stability region is smaller than that of the TO policy, at $E[X]$ close to $E[g(Y)]$, the Greedy performance rapidly deteriorates. Thus it is worthwhile to look for some other good policy. Notice that the TO policy wastes energy if $q_k < g(E[Y] - \epsilon)$. Therfore, we can improve upon it by saving the energy $(E[Y]-\epsilon - g^{-1}(q_k))$ and using it when the $q_k$ is greater than $g(E[Y]-\epsilon)$. However for $g$ a log function, using a large amount of energy $t$ is also wasteful even when $q_k > g(t)$. Taking into account these facts we improve over the TO policy as \begin{equation} T_k = min(g^{-1}(q_k), E_k, 0.99 (E[Y] + 0.001 (E_k - cq_k)^+)) \label{onestar} \end{equation} where $c$ is a positive constant. The improvement over the TO also comes from the fact that if $E_k$ is large, we allow $T_k > E[Y]$ but only if $q_k$ is not very large. The constants 0.99 and 0.001 were chosen by trial and error from simulations after experimenting with different scenarios.
 
We will see in Section \ref{simulation} via simulations that this policy, to be denoted by MTO can indeed provide lower mean delays than TO at loads above $E[g(Y)]$. 

One advantage of (\ref{disp}) over (\ref{eqn5}) and (\ref{onestar}) is that while using (\ref{disp}), after some time $T_k =  E[Y]-\epsilon$. Also, at any time, either one uses up all the energy or uses $E[Y]-\epsilon$. Thus one can use this policy even if exact information about $E_k$ is not available (measuring $E_k$ may be difficult in practice). In fact, (\ref{disp}) does not need even $q_k$ while (\ref{eqn5}) either uses up all the energy or uses $f(q_k)$ and hence needs only $q_k$ exactly.

Now we show that under the greedy policy (\ref{eqn5}) the queueing process is stable when $E[X] < E[g(Y)]$. In next few results (Theorems 2 and 3) we assume that the energy buffer is finite, although large. For this case Lemma 1 and Theorem 1 also hold under the same assumptions with slight modifications in their proofs.

\vspace{0.1cm}

\noindent
\textbf{Theorem 2} If the energy buffer is finite, i.e., $E_k \leq \bar{e} < \infty$  (but $\bar{e}$ is large enough) and $E[X] < E[g(Y)] $ then under the greedy policy (\ref{eqn5}), ${ (q_k, E_k)}$ has an Ergodic set.  \hspace*{1.5cm} $ \blacksquare$

The above result will ensure that the Markov chain $\{(q_k, E_k) \}$ is ergodic and hence has a unique stationary distribution if $\{(q_k, E_k) \} $ is irreducible. A sufficient condition for this is $0 < P[X_k=0] < 1$ and $0 < P[Y_k=0] < 1$ because then the state $(0,0)$ can be reached from any state with a positive probability. In general, $\{(q_k, E_k) \}$ can have multiple ergodic sets. Then, depending on the initial state, $\{(q_k, E_k) \}$ will converge to one of the ergodic sets and the limiting distribution depends on the initial conditions.

\section{Optimal Policies}
\label{opt}
In this section we choose $T_k$ at time $k$ as a function of $q_k$ and $E_k$ such that  \begin{equation*}
E\left[ \sum_{k=0}^{\infty} \alpha^k \ q_k \right] 
\end{equation*} is minimized where $0 < \alpha < 1$ is a suitable constant. The minimizing policy is called $\alpha$-discount optimal. When $\alpha = 1$, we minimize \begin{equation*} \lim_{n \rightarrow \infty}\sup \\\frac{1}{n} E \left[ \sum^{n-1}_{k=0} q_k \right]. \end{equation*} This optimizing policy is called average cost optimal. By Little's law (\cite{asmussen}) an average cost optimal policy also minimizes mean delay. If for a  given $(q_k, E_k)$, the optimal policy $T_k$ does not depend on the past values, and is time invariant, it is called a stationary Markov policy.

If $\{ X_k \}$ and $\{ Y_k \}$ are Markov chains then these optimization problems are Markov Decision Problems (MDP). For simplicity, in the following we consider these problems when $\{ X_k \}$ and $\{ Y_k \}$ are \emph{iid}. We obtain the existence of optimal $\alpha$-discount and average cost stationary Markov policies. 

\vspace{0.1cm}

\noindent
\textbf{Theorem 3} If $g$ is continuous and the energy buffer is finite, i.e., $e_k \leq  \bar{e} < \infty$ then there exists an optimal  $\alpha$-discounted  Markov stationary policy. If in addition $E[X] < g(E[Y])$ and $E[X^2] < \infty$, then there exists an average cost optimal stationary Markov policy. The optimal cost $v$ does not depend on the initial state. Also, then the optimal $\alpha$-discount policies tend to an optimal average cost policy as $\alpha \rightarrow 1$. Furthermore, if $v_\alpha (q,e) $ is the optimal $\alpha$-discount cost for the initial state $(q,e)$ then 
\begin{equation*} 
\lim_{\alpha \rightarrow 1}(1- \alpha) \ \mbox{inf}_{(q,e)} v_\alpha (q,e) = v   \end{equation*} \hspace*{8.0cm} $ \blacksquare$

In Section \ref{stability} we identified a throughput optimal policy when $g$ is nondecreasing, concave. Theorem 3 guarantees the existence of an optimal mean delay policy. It is of interest to identify one such policy also. In general one can compute an optimal policy numerically via Value Iteration or Policy Iteration but that can be computationally intensive (especially for large data and energy buffer sizes). Also it does not provide any insight and requires traffic and energy profile statistics. In Section \ref{stability} we also provided a greedy policy (\ref{eqn5}) which is very intuitive,  and is throughput optimal for linear $g$. However for concave $g$ (including the cost function $\frac{1}{2} log (1+\gamma t))$ it is \emph{not} throughput optimal and provides low mean delays only for low load. Next we show that it provides minimum mean delay for linear $g$. 

\noindent

\textbf{Theorem 4} The Greedy policy (\ref{eqn5}) is $\alpha$-discount optimal for $ 0 < \alpha < 1$ when $g(t) = \gamma t$ for some $\gamma >0$. It is also average cost optimal.  \hspace*{5.5cm} $ \blacksquare$

The fact that Greedy is  $\alpha$-discount optimal as well as average cost optimal implies that it is good not only for long term average delay but also for transient mean delays.
 
\section{Generalizations}
\label{general}

In this section we consider two generalizations. First we will extend the results to the case of fading channels and then to the case where the sensing and the processing energy at a sensor node are non-negligible with respect to the transmission energy.

In case of fading channels, we assume flat fading during a slot. In slot $k$ the channel gain is $h_k$. The sequence $\{ h_k \}$ is assumed stationary, ergodic, independent of the traffic sequence $\{ X_k \}$ and the energy generation sequence $\{ Y_k \}$. Then if $T_k$ energy is spent in transmission in slot $k$, the $\{ q_k \}$ process evolves as \begin{equation*} 
q_{k+1} = (q_k -g(h_k T_k))^+ + X_k.
\end{equation*} If the channel state information (CSI) is not known to the sensor node, then $T_k$ will depend only on $(q_k, E_k)$. One can then consider the policies used above. For example we could use $T_k = min(E_k, E[Y]-\epsilon)$. Then the data queue is stable if $E[X] < E[g(h(E[Y]-\epsilon))]$. We will call this policy unfaded TO. If we use Greedy (\ref{eqn5}), then the data queue is stable if $E[X] < E[g(hY)]$.

If CSI $h_k$ is available to the node at time $k$, then the following are the throughput optimal policies. If $g$ is linear, then $g(x) = \beta x$ for some $\beta > 0$. Then, if $ 0 \leq h \leq \bar{h} < \infty$ and $P(h = \bar{h}) > 0$, the optimal policy is:  $T(\bar{h})=(E[Y]-\epsilon)/p(h=\bar{h})$ and $T(h) = 0$ otherwise. Thus if $h$ can take an arbitrarily large value with positive probability, then $E[hT(h)] = \infty$ at the optimal solution. 

If $g(x) = \frac{1}{2} log(1+ \beta x)$, then the water filling (WF) policy
\begin{equation}
T_k (h) = \left(\frac{1}{h_0}-\frac{1}{h}\right)^+\label{fto} 
\end{equation} where $h_0$ is obtained from the average power constraint $E[T_k] = E[Y] - \epsilon$, is throughput optimal. This is because it maximizes $\frac{1}{2} E_h[log (1+ \beta h T(h))]$ with the given constraints. 

Both of the above policies can be improved as before, by not wasting energy when there is not enough data. As in (\ref{onestar}) in Section \ref{stability}, we can further improve WF by taking \begin{equation} T_k = min\left(f(q_k), E_k, \left(\frac{1}{h_0}-\frac{1}{h} + 0.001 (E_k - cq_k)^+\right)^+\right) \label{twostar}.\end{equation} We will call it MWF.
These policies will not minimize mean delay. For that, we can use the MDP framework used in Section \ref{opt} and numerically compute the optimal policies.

Till now we assumed that all the energy that a node consumes is for transmission. However, sensing, processing and receiving (from other nodes) also require significant energy, especially in more recent higher end sensor nodes (\cite{raghunathan}). Since we have been considering a single node so far, we will now include the energy consumed by sensing and processing only. For simplicity, we will assume that the node is always in one energy mode (e.g., lower energy modes (\cite{sinha}) available for sensor nodes will not be considered). If a sensor node with an energy harvesting system can be operated in energy neutral operation in normal mode itself (i.e., it satisfies the conditions in Lemma 1), then there is no need to have lower energy modes. Otherwise one has to resort to energy saving modes.

We will assume that $Z_k$ is the energy consumed by the node for sensing and processing in slot $k$. Unlike $T_k$ (which can vary according to $q_k$), $ \{ Z_k \}$ can be considered a stationary, ergodic sequence. The rest of the system is as in Section \ref{model}. Now we briefly describe a energy management policy which is an extension of the TO policy in Section \ref{stability}. This can provide an energy neutral operation in the present case. Improved/optimal policies can be obtained for this system also but will not be discussed due to lack of space.

Let $c$ be the minimum positive constant such that $E[X] < g(c)$. Then if $ c + E[Z] < E[Y]-\delta $, (where $\delta$ is a small positive constant) the system can be operated in energy neutral operation: If we take $ T_k \equiv c$ (which can be done with high probability for all $k$ large enough), the process $ \{ q_k \}$ will have a unique stationary, ergodic distribution and there will always be energy $Z_k$ for sensing and processing for all $k$ large enough.  The result holds if $\{ (X_k, Y_k, Z_k) \}$ is an ergodic stationary sequence. The arguments to show this are similar to those in Section \ref{stability} and are omitted.

When the channel has fading, we need $E[X] < E[g(ch)] $ in the above paragraph.

\section{Simulations for single node}
\label{simulation}

In this section, we compare the different policies we have studied via simulations. The $g$ function is taken as linear $(g(x) = 10x)$ or as  $g(x) = log (1+x)$ . The sequences $\{ X_k \}$ and $\{ Y_k \}$ are \emph{iid}. (We have also done limited simulations when $\{ X_k \}$ and $\{ Y_k \}$ are Autoregressive and found that conclusions drawn in this section continue to hold). We consider the cases when $X$ and $Y$ have truncated Poisson, exponential, Erlang or Hyperexponential distributions. The policies considered are: Greedy, TO, $T_k \equiv Y_k$, MTO (with $c=0.1$) and the mean delay optimal. At the end, we will also consider channels with fading. For fading channels we compare unfaded TO and MTO against fading TO and fading MTO. For the linear $g$, we already know that the Greedy policy is throughput optimal as well as mean delay optimal. 

\begin{figure}[!ht]
\begin{center}
\includegraphics[scale=0.3]{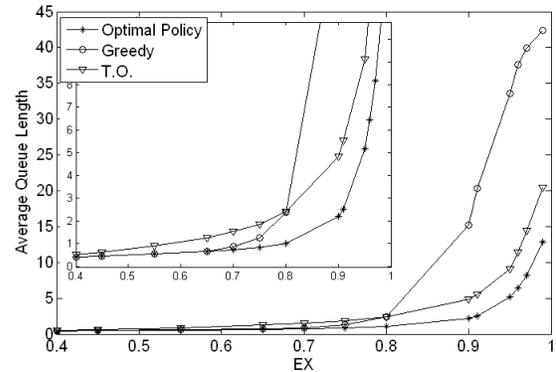}

\caption{Mean Delay Optimal, Greedy, TO Policies with No Fading; Nonlinear $g$; Finite, Quantized data and energy buffers; $X,Y$: Poisson truncated at 5; $E[Y]=1,E[g(Y)]=0.92,g(E[Y])=1$}

\label{plot1}

\end{center}
\end{figure}
\begin{figure}[!ht]
\begin{center}
\includegraphics[scale=0.3]{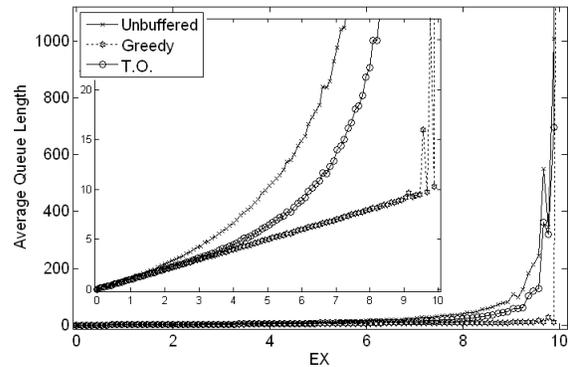}

\caption{Comparison of policies with No Fading; $g(x)=10x$; $X,Y$: Exponential; $E[Y]=1,E[g(Y)]=10,g(E[Y])=10$}

\label{plot2}

\end{center}
\end{figure}

\begin{figure}[!ht]
\begin{center}
\includegraphics[scale=0.3]{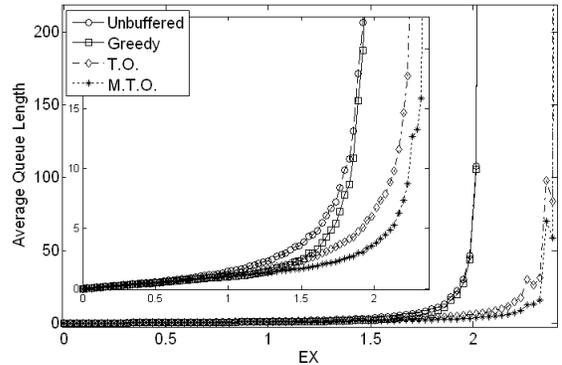}

\caption{Comparison of policies with No Fading; $g(x)=log(1+x)$; $X,Y$: Exponential; $E[Y]=10,E[g(Y)]=2.01,g(E[Y])=2.4$}
 
\label{plot4}

\end{center}
\end{figure}

\begin{figure}[!ht]
\begin{center}
\includegraphics[scale=0.3]{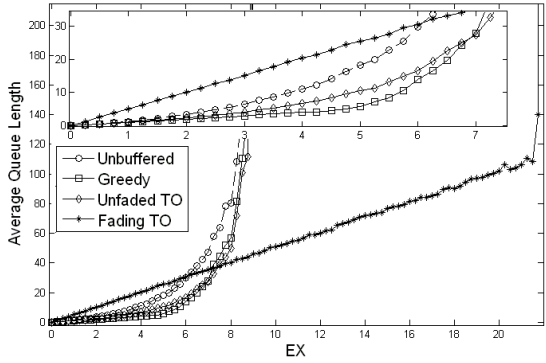}
\caption{Comparison of policies with Fading; $g(x)=10x$; $X,Y$: Hyperexponential(5); $E[Y]=1,E[g(Y)]=10,g(E[Y])=10$}

\label{plot7}

\end{center}
\end{figure}  

\begin{figure}[!ht]
\begin{center}
\includegraphics[scale=0.3]{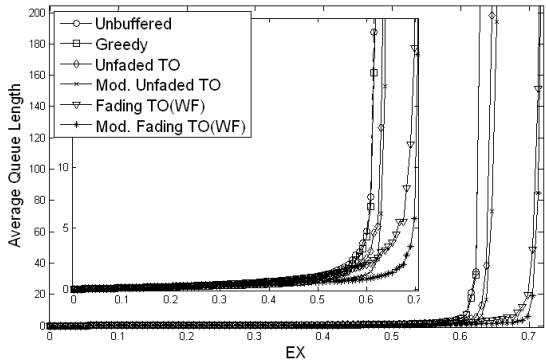}

\caption{Comparison of policies with Fading; $g(x)=log(1+x)$; $X,Y$: Erlang(5); $E[Y]=1,E[g(hY)]=0.62,E[g(hE[Y])]=0.64$; WF, Mod. WF stable for $E[X]<0.70$}

\label{plot8}
\end{center}
\end{figure}

The mean queue lengths for the different cases are plotted in Figs. \ref{plot1}-\ref{plot8}. 

In Fig. \ref{plot1}, we compare Greedy, TO and mean-delay optimal (OP) policies for nonlinear $g$. The OP was computed via Policy Iteration. For numerical computations, all quantities need to be finite. So we took data and energy buffer sizes to be $50$ and used quantized versions of $q_k$ and $E_k$. The distribution of $X$ and $Y$ is Poisson truncated at $5$. These changes were made only for this example. Now $g(E[Y]) = 1$ and $E[g(Y)] = 0.92$. We see that the mean queue length of the three policies are negligible till $E[X]=0.8$. After that, the mean queue length of the Greedy policy rapidly increases while performances of the other two policies are comparable till $1$ (although from $E[X] = 0.6$ till close to $1$, mean queue length of TO is approximately double of OP). At low loads, Greedy has less mean queue length than TO. 

Fig. \ref{plot2} considers the case when $X$ and $Y$ are exponential and $g$ is linear. Now $ E[Y] = 1$ and $g(E[Y]) = E[g(Y)] = 10.$ Now all the policies considered are throughput optimal but their delay performances \emph{differ}. We observe that the policy $T_k \equiv Y_k$ (henceforth called unbuffered) has the worst performance. Next is the TO. 


Fig. \ref{plot4} provides the above results for $g$ nonlinear, when $X$ and $Y$ are exponential. Now, as before, $T_k \equiv Y_k$ is the worst. The Greedy performs better than the other policies for low values of $E[X]$. But Greedy becomes unstable at $E[g(Y)]=2.01$  while the throughput optimal policies become unstable at $g(E[Y]) = 2.40$. Now for higher values of $E[X]$, the modified TO performs the best and is close to Greedy at low $E[X]$. 

Figs. \ref{plot7}-\ref{plot8} provide results for fading channels. The fading process $\{ h_k \}$ is \emph{iid} taking values $0.1, 0.5, 1.0$ and $2.2$ with probabilities $0.1, 0.3, 0.4$ and $0.2$ respectively. Fig. \ref{plot7} is for linear $g$ and Fig. \ref{plot8} is for nonlinear $g$. The policies compared are unbuffered, Greedy, Unfaded TO (\ref{disp}) and Fading TO. In Fig. \ref{plot8}, we have also considered Modified Unfaded TO (\ref{onestar}) and Modified Fading TO (MWF) with $c=0.1$ (\ref{twostar}).

In Fig. \ref{plot7}, $X$ and $Y$ have Hyperexponential distributions. The distribution of r.v. $X$ is a mixture of 5 exponential distributions with means $E[X]/4.9, 2E[X]/4.9, 3E[X]/4.9, 6E[X]/4.9$ and $10E[X]/4.9$ and probabilities $0.1, 0.2, 0.2, 0.3$ and $0.2$ respectively. The distribution of $Y$ is obtained in the same way. Now $E[Y] = 1$, $E[g(hY)]=10$ and $E[g(hE[Y])]=10$.  The stability region of fading TO is $E[X]<E[g(\bar{h}Y)] =22.0$ while that of the other three policies  is $E[X]<10$. However, mean queue length of fading TO is larger from the beginning till almost 10. This is because in fading TO, we transmit only when $h=\bar{h}=2.2$ which has a small probability ($=0.2$).

Fig. \ref{plot8} considers nonlinear $g$ with $X,Y$ Erlang distributed. Also, $E[Y]=1, E[g(hY)]=0.62, E[g(hE[Y])] = 0.64$.  Now we see that the stability region of unbuffered and Greedy is the smallest, then of TO and MTO while WF and MWF provide the largest region and are stable for $E[X]<0.70$. MTO and MWF provide improvements in mean queue lengths over TO and WF. 

\section{Multiple Access Channel}
\label{tree}
In a sensor network, all the nodes need to transmit their data to a fusion node. Thus, for this a natural network to consider is a Tree (\cite{baek}). In the present scenario of nodes with energy harvesting sources, selection of a Tree can depend on the energy profiles of different nodes. This will be subject of another study. Here we assume that an appropriate Tree has been formed and will concentrate on the link layer. 

An important building block for this network is a multiple access channel. In sensor networks contention based (e.g., CSMA) and contention-free (TDMA/CDMA/FDMA) MAC protocols are considered suitable (\cite{akyildiz}, \cite{kredo}). In fact for estimation of a random field, contention-free protocols are more appropriate.  

We consider the case where $N$ nodes with data queues $Q_1, ..., Q_N$ are sharing a wireless channel. Each queue generates its traffic, stores in a queue and transmits as in Section \ref{model}. Also, each node has its own energy harvesting mechanism. The traffic generated at different queues and their energy mechanisms are assumed independent of each other.

Let $\{ X_k(i) \}, \{Y_k(i) \}$ and $\{ Z_k(i) \}$ be the sequences corresponding to node $i$. For simplicity we will assume $\{X_k(i), k \geq 0 \}$ and $\{ Y_k(i), k \geq 0 \}$ to be \emph{iid} although these assumptions can be weakened as for a single queue. As mentioned at the end of Section \ref{general}, the energy consumption $\{ Z_k(i) \}$ can be taken care of if we simply replace $E[Y(i)]$ by $E[Y(i)] - E[Z(i)]$ in our algorithms. In the following we  do that and write it as $E[Y(i)]$ only (and hence ignore $Z_k(i)$).

The $N$ queues can share the channel in different ways. The stability region of $Q_1, Q_2,..., Q_N$ and optimal (good) transmit policies depend upon the sharing mechanism used. We consider a few commonly used scenarios in the rest of the paper. 

\section{Orthogonal Channels}
\label{orthogonal}

The $N$ sensor nodes use TDMA/orthogonal CDMA/FDMA/OFDMA to transmit. Then the $N$ queues become independent, decoupled queues and can be considered separately. Thus, the transmission policies developed in previous sections for a single queue can be used here.  In the following we explain them in the context of TDMA.  

If the queues have to use the channel in a TDMA fashion then necessary conditions for stability of the $N$ queues are : There exist $\alpha_1, \alpha_2,.... \alpha_N, \alpha_i \geq 0$ and $\sum_{i=1}^{N} \alpha_i = 1 $ such that
\begin{equation}
 E[X(i)] < \alpha_i \ g_i \ \left( \frac{E[Y(i)]}{\alpha_i} \right) ,\ \ i=1,2,..., N,  \label{meqn6}
\end{equation} where $g_i$ is the energy to bit mapping for $Q_i$.  A stable policy for each queue will be as in Section \ref{stability}: $Q_i$ is given $\alpha_i$ fraction of slots (on a long term basis) and it uses energy $(E[Y(i)] - \epsilon)/ \alpha_i$ whenever it transmits. For better delay performance, the slots allocated to different queues should be uniformly spaced. We can improve on the mean delay by using (\ref{onestar}).

It is possible that more than one set of $(\alpha_1, ..., \alpha_N)$ satisfy the stability condition (\ref{meqn6}). Then one should select the values which minimize a cost function, (say) weighted sum of mean delays.

\section{Opportunistic Scheduling for fading channels: Orthogonal Channels}
\label{opportunistic}

Now we discuss the MAC with fading. Let $\{ h_k(i), \ k \geq 1 \}$ be the channel gain process for $Q_i$. It is assumed stationary, ergodic and independent of the fading process for $Q_j, j \neq i$. We discuss opportunistic scheduling for the contention free MAC. We will study the CSMA based algorithms in the next Section.

If we assume that each of $Q_i$ has infinite data backlog, then the policy 
that maximizes the sum of throughputs for $g(x) = log(1+ \beta x)$ and for symmetric statistics (i.e., each $h_i$ has same statistics and all $E[Y(i)]$ are same) is to choose $Q$ \begin{equation} i^*_k = arg \ max(h_k(i)) \label{meqn10}
\end{equation} in slot $k$ and use $T_k(h)$ via the water-filling formula (\ref{fto}) with the average power constraint \begin{equation*} E[T_k(h)] = N  E[Y(i^*_k)-\epsilon]. \end{equation*} This is an extension of an algorithm in \cite{knopp} to the energy harvesting nodes. A modification of this policy is available in \cite{knopp} for asymmetric case.

If $g$ is linear, then for the symmetric case, a channel is selected only if it has the highest possible gain (for $h$ bounded). If more than one channel is in the best state, select one of them with equal probability.

Although (\ref{meqn10}) maximizes throughput, it may be unfair to different queues and may not provide the QoS. Furthermore, in our setup infinite backlog is not a realistic assumption. Without this assumption a throughput optimal policy (in the class of policies which use constant powers) is to choose queue
\begin{equation} \label{meqn11}
i^*_k = arg \ max \left( q_k(i) g_i \left( h_k(i) \left( \frac{E[Y(i)]-\epsilon}{\alpha(i)} \right) \right) \right) 
\end{equation} and then use $T_k = (E[Y(i^*_k)] - \epsilon) / \alpha(i^*_k)$. Here $\alpha(i^*_k)$ is the fraction of time slots assigned to $i^*_k$. However now we do not know $\alpha(i^*_k)$ and this may be estimated (see the end of this section).  If $\alpha(i^*_k)$ is replaced with the true value, then stability of the queues in the MAC follows from \cite{eryilmaz} if the fading states take values in a finite set and the system satisfies the following condition. Let there exist a function $f(r_k(1), ..., r_k(N))$ which picks one of the queues as a function of $ (r_k(1), ..., r_k(N))$ where $r_k(i) = g_i \left( h_k(i) \frac{E[Y(i)] - \epsilon}{\alpha_i} \right)$, $\alpha(i) \triangleq E_{\pi} [1 \{ f(r_1, ..., r_N) = i \}]$ and $\pi (r_1, ..., r_N)$ is the stationary distribution of $(r_1, ..., r_N)$.  Then if $E[X(i)] < \sum r_i 1 \{ f(r_1, ..., r_N)=i \} \pi (r_1, ..., r_N)$ for each $i$, the system is stable. This policy tries to satisfy the traffic requirements of different nodes but may not be delay optimal. Based on experience in \cite{sharma}, a Greedy policy 
\begin{equation} \label{meqn12}
i^*_k = arg \ max \left( min \left(g_i \left( h_k(i) \left( \frac{E[Y(i)]-\epsilon}{\alpha(i)} \right),q_k(i) \right) \right) \right)
\end{equation} provides better mean delays. However, it is throughput optimal only for symmetric traffic statistics and when $E[Y(i)] = E[Y(j)],$ for all $ i,j$. But it can be made throughput optimal (as (\ref{meqn11}))  while still retaining (partially) its mean delay performance as follows. Choose an appropriately large positive constant $L$. If none of $q_k(i)$ is greater than $L$, use (\ref{meqn12}); otherwise, on the set $\{ i: \ q_k(i) > L, \}$, use (\ref{meqn11}). We call this Modified Greedy Policy.

The mean delay of the above policies can be further improved, if instead of $T_k = (E[Y(i)] - \epsilon) / \alpha(i)$, we use (\ref{onestar}). But the stability region remains same.

The policies (\ref{meqn11}) and (\ref{meqn12}) can be further improved if instead of using $T_k(i) = E[Y(i)] - \epsilon$, we use waterfilling for $g$ in (\ref{allert3}). Of course we reduce transmit power as in (\ref{onestar}) if there is not enough data to transmit. Now not only the mean delays reduce but the stability region also enlarges.

The policies (\ref{meqn11}), (\ref{meqn12}) and Modified Greedy provide good performance, require minimal information (only $E[Y(i)]$), are easy to implement and have low computational requirements. In addition they naturally adapt to changing traffic and channel conditions.

In (\ref{meqn11})-(\ref{meqn12}) we need $\alpha(i)$ to obtain the energy $T_k$. But unlike for TDMA, $\alpha(i_k)$ is not available in these algorithms and depends on the algorithm used. Thus, in these algorithms we use a simple variant of the LMS (Least Mean Square) algorithm (\cite{haykin}) to estimate $\alpha(i)$:

\vspace{0.1in}
\noindent
Initially start with guess
\begin{equation*}
\alpha_0(i) = \frac{1}{N} , \ i=1, ..., N.
\end{equation*} Run the algorithm for (say) $L_1$ number of slots. Each node $i$ computes the fraction $\alpha^{'}(i)$ of slots it gets and recomputes 
\begin{equation} \label{meqn13}
\alpha_{n+1}(i) = \alpha_n (i) - \mu (\alpha_n(i) - \alpha^{'}(i))
\end{equation} where $\mu$ is a small postive constant. 

At any time the current estimate of $\alpha(i)$ is used by the algorithms.

\section{Opportunistic scheduling for fading channels: CSMA}
\label{csma}

Since ZigBee and 802.11 use CSMA, we discuss opportunistic scheduling for CSMA also. As against (\ref{meqn11})-(\ref{meqn12}), this is a completely decentralized algorithm. This is used by Zhao and Tong \cite{zhao}. The basic idea in \cite{zhao} is to make the back-off mechanism in a node to be a function of the channel state of that node. The nodes that are to be given priority are given smaller back-off time. This mechanism has also been used in IEEE 802.11e to provide priority to voice and video traffic. In this section we take this idea further by also including the effect of queue lengths and power control in deciding the back-off interval as against only the channel state in \cite{zhao}. 

Let $f$ be a nonincreasing function with values in $[0,\tau_{max}]$ where $\tau_{max}$ is the maximum allowed back-off time in slots. If $h$ is the channel gain in a slot then in \cite{zhao} the back-off time is taken to be $f(h)$. 

In our setup, to use opportunistic scheduling in CSMA, we use the above mentioned monotonic function $f$ on each of the sensor nodes contending for the channel. The $Q_i$ uses the back off time of 
\begin{equation} \label{meqn14}
f \left( g_i \left( h_k(i)\frac{(E[Y(i)]-\epsilon)}{\alpha(i)} \right) \right).
\end{equation} When a node gets the channel, it will transmit a complete packet and use energy per slot as 
\begin{equation} \label{meqn15}
T_k(i^*_k) = [E[Y(i^*_k)] -\epsilon] / \alpha(i^*_k).
\end{equation} Now we are making the usual assumption that the channel gains stay constant during the transmission of a packet. We can use (\ref{onestar}) to improve performance.

Using the ideas in the last section we can develop better algorithms than (\ref{meqn14})-(\ref{meqn15}). Indeed, with (\ref{meqn14}), instead of (\ref{meqn15}), we can use waterfilling (for $g$ in (\ref{allert3})). We can also improve over (\ref{meqn14}) by using, for back-off time of $i$th node,
\begin{equation} \label{meqn16}
f \left( q_k(i) \ g_i \left( h_k(i) \frac{E[Y(i)] - \epsilon}{\alpha(i)} \right) \right) 
\end{equation} which takes care of the traffic requirements of different nodes. 

We can also use the (modified) Greedy in (\ref{meqn12}). The $\alpha(i)$ in the above algorithms will be computed via LMS in (\ref{meqn13}).

We will compare the performance of these algorithms via simulations in Section \ref{msimulation}.

An advantage of above algorithms over the algorithms in Section \ref{opportunistic} are that these are completely decentralized: Each node uses only its own queue length, channel state and $E[Y(i)]$ to decide when to transmit. The algorithms in Section \ref{opportunistic} require a central controller (may be a cluster head) for implementation. Centralized algorithms have also been considered in sensor networks and provide better performance.
  
\section{Simulations for MAC Protocols}
\label{msimulation}

In this section for simplicity, we simulate the system under symmetric conditions, apply the different algorithms, and compare their performances. We use $g(x) = log(1+x)$. The fading of each channel changes from slot to slot independently; $\{ h_k(i), k \geq 1 \}$ are \emph{iid} with values 0.1, 0.6, 1.8 and 5 with probabilities 4/12, 5/12, 2/12 and 1/12. The $\{ X_k(i), k \geq 1 \}$ and $\{ Y_k(i), k \geq 1 \}$  are iid expontential. The LMS (\ref{meqn13}) was taken with $\mu=0.01$ and the $\alpha_k$s were updated after 30-50 slots.

\begin{figure}[!ht]
\begin{center}
\includegraphics[scale=0.4]{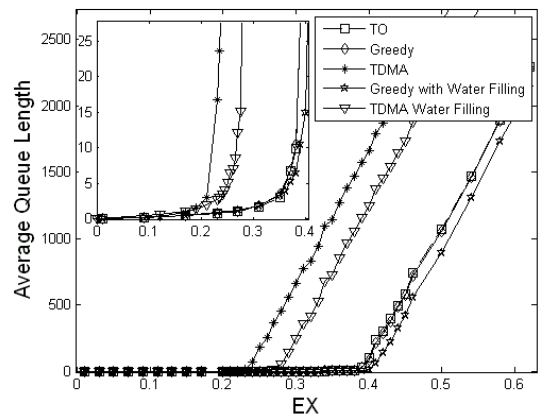}

\caption{Orthogonal Channels: Symmetric, 3 Queues} \label{orth}
\end{center}
\end{figure}

\begin{figure}[!ht]
\begin{center}
\includegraphics[scale=0.4]{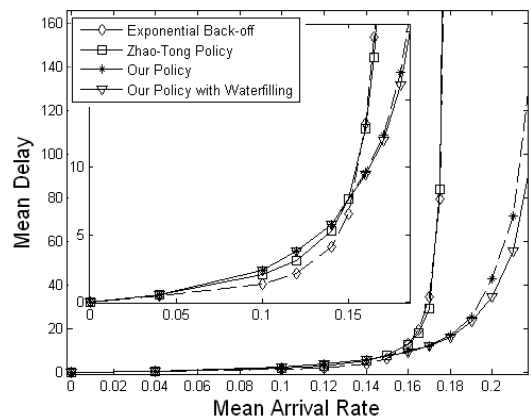}

\caption{CSMA: Mean Delay, Symmetric 10 Queues} \label{delay}
\end{center}
\end{figure}
\begin{figure}[!ht]
\begin{center}
\includegraphics[scale=0.4]{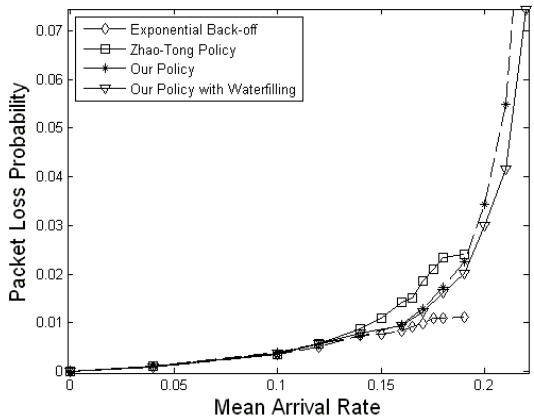}

\caption{CSMA: Packet Loss Probability, Symmetric 10 Queues} \label{pl}
\end{center}
\end{figure}

For orthogonal channels, under symmetric conditions with 3 queues, average queue lengths are shown in Fig. \ref{orth} for TO (\ref{meqn11}), Greedy (\ref{meqn12}), TDMA, Greedy with water-filling and TDMA with water-filling policies. The $\{ h_k(i), k \geq 1 \}$ are \emph{iid} exponential with mean 1. For symmetric conditions, Greedy is throughput optimal and hence Modified Greedy is not implemented. We see that TDMA becomes unstable much before the other policies, and that its average queue length is much worse even when it is stable. Greedy performs better than TO near the stability boundary which is $E[X]=0.39$. Water-filling improves the stability region of TDMA as well as Greedy.

For CSMA, Figs. \ref{delay} and \ref{pl} show mean delays and packet loss probabilities under symmetric conditions with 10 queues and with normal exponential backoff, Zhao-Tong \cite{zhao}, our policy (\ref{meqn16}) and our policy with water-filling (with $f_{policy}(x) = \beta_{policy}/x$ and $Ef=1.55$ at $EX=0.17; h$ assumes values 0.1,0.5,1.0,2.2 for time fractions 0.1,0.3,0.4,0.2). We simulated the 10 queues in continuous time. Also, $E[Y]=1$ and the data packets of unit size arrive at each queue as Poisson streams. We see that opportunistic policies improve mean delays substantially.

\section{Conclusions}
\label{conclude}
We have considered sensor nodes with energy harvesting sources, deployed for random field estimation. Throughput optimal and mean delay optimal energy management policies for single nodes are identified which can make them work in energy neutral operation. Next these results are extended to fading channels and when energy at the sensor node is also consumed in sensing and data processing. Similarly we can include leakage/wastage of energy when it is stored in the energy buffer and when it is extracted. Finally these policies are used to develop efficient MAC protocols for such nodes. In particular versions of TDMA, opportunistic MACs for fading channels and CSMA are developed. Their performance is compared via simulations. It is shown that opportunistic policies can substantially improve the performance.


\end{document}